\documentclass[a4paper]{article}

\usepackage{INTERSPEECH2022}
\usepackage{hyperref}
\usepackage{makecell}
\usepackage{multirow}
\usepackage{booktabs}
\usepackage{xcolor}
\usepackage{soul}
\usepackage{adjustbox}
\usepackage{graphicx}

\usepackage[natbib, bibencoding=utf8, citestyle=numeric, bibstyle=ieee, maxbibnames=999, maxcitenames=2, mincitenames=1, sortcites]{biblatex}
\bibliography{main}

\usepackage{cleveref}

\usepackage[activate]{microtype}
\newcommand{\eg}{e.\,g.,\,}
\newcommand{\ie}{i.\,e.,\,}

\newcommand{\et}{{et al.\,}}

\title{A Temporal-oriented Broadcast ResNet for COVID-19 Detection}
\name{Xin Jing$^1$, Shuo Liu$^1$, Emilia Parada-Cabaleiro$^2$, Andreas Triantafyllopoulos$^1$,\\Meishu Song$^1$, Zijiang Yang$^1$, Bj\"orn W.\ Schuller$^{1,3}$}
\address{
  $^1$Chair of Embedded Intelligence for Health Care \& Wellbeing, University of Augsburg, Germany\\
  $^2$ Institute of Computational Perception, Johannes Kepler University Linz, Austria\\
  $^3$GLAM -- Group on Language, Audio, \& Music, Imperial College London, UK}
\email{xin.jing@informatik.uni-augsburg.de}

\begin{document}

\maketitle
\begin{abstract}
Detecting COVID-19 from audio signals, such as breathing and coughing, can be used as a fast and efficient pre-testing method to reduce the virus transmission. Due to the promising results of deep learning networks in modelling time sequences, and since applications to rapidly identify COVID in-the-wild should require low computational effort, we present a temporal-oriented broadcasting residual learning method that achieves efficient computation and high accuracy with a small model size. Based on the EfficientNet architecture, our novel network, named Temporal-oriented ResNet~(TorNet), constitutes of a broadcasting learning block, \ie the Alternating Broadcast (AB) Block, which contains several Broadcast Residual Blocks (BC ResBlocks) and a convolution layer. With the AB Block, the network obtains useful audio-temporal features and higher level embeddings effectively with much less computation than Recurrent Neural Networks~(RNNs), typically used to model temporal information. TorNet achieves 72.2\% Unweighted Average Recall (UAR) on the INTERPSEECH 2021 Computational Paralinguistics Challenge COVID-19 cough Sub-Challenge, by this showing competitive results with a higher computational efficiency than other state-of-the-art alternatives. 

\end{abstract}

\section{Introduction}
COVID-19 cases are still rising in several countries, indicating that the pandemic is still a main health challenge for our world\cite{who2022covid}. Although there are rapid testing methods, their efficiency is often limited by the capacity of the testing equipment. In addition, as their production depends on the materials' availability, limited resources might yield crowds that in turn, paradoxically, increase the infection rates. Indeed, ubiquitous \textit{low-cost} methods for detecting COVID-19 are still being explored. In the realm of Artificial Intelligence, Deep Neural Networks~(DNNs)~have been growing in popularity in recent years, setting the state-of-art in a variety of tasks, including COVID-19 detection from audio signals, \eg patients' breathing and coughing~\cite{schuller2021covid,andrew2021}.

The temporal component is an essential characteristic of audio signals. Thus, learning discriminated representations containing temporal information is crucial to achieve a better classification network when working with audio~\cite{deshpande2022ai,schuller2021covid}. To make full use of temporal information, Recurrent Neural Networks~(RNNs)~and variants with Long-Short Term Memory~(LSTM)~\cite{baird2021emotion,yan21c_interspeech,alvarez2021motivic} have been successfully developed. However, RNNs are computationally more intensive and require more storage compared to a typical Convolutional Neural Network~(CNN). Previous works have also shown that transformers can exploit the temporal properties of audio to obtain higher detection results than RNNs~\cite{gong2021ast, Song22-PCA,chang2021}. Still, over-parametrised transformer-based deep networks might be prone to overfitting, and similar to CNNs, computationally inefficient. 

The successful application of RNNs and transformers to audio data illustrates the importance of temporal features for audio tasks. Nevertheless, the complexity of these network structures, unlike CNNs, increases the computational complexity and reduces the training stability. In the present work, we propose a temporal broadcast residual convolution block, \ie the Alternating Broadcast Block (AB Block), in which we average the 2D features in the frequency dimension to guide the network's focus on the temporal features. Inspired by the EfficientNet~\cite{tan2019efficientnet} architecture (made up of repeated blocks and based on the residual learning), we introduce a new deep learning network named Temporal-oriented ResNet~(TorNet) that contains several AB Blocks to make full use of the temporal information in the audio segments. Furthermore, we also adopt Instance Normalisation~\cite{ulyanovVL16} to assist the network to find the relevant feature areas of the Mel-spectrogram, by this improving the classification results. We evaluate the efficiency of TorNet on the detection of COVID-19 from coughing signals, using the audio dataset from the INTERSPEECH 2021 Computational Paralinguistics Challenge's COVID-19 cough sub-challenge~\cite{schuller2021compare}. For reproducibility purposes, the source code of our work is freely available\footnote{https://github.com/EIHW/compare21\_tornet\_covid}.

The remainder of our paper is organised as follows: We summarise the related research in Section  \ref{sec:Related_works}. Then, we present our network architecture and describe the experimental settings in Sections \ref{sec:model} and \ref{sec:Experiments}, respectively. In Section \ref{sec:results}, we discuss the results. Finally, in Section  \ref{sec:conclusion}, we conclude with a brief summary and outline future directions.  

\section{Related Works}
\label{sec:Related_works}
Data representations such as Mel-Spectrograms can be seen from two different perspectives: either as an image, or as an audio sequence. This duality leads to the use of a variety of DNN architectures typical of both Computer Vision~(CV) and the audio domain~\cite{purwins2019deep,yang2020learning}.

On the one side, previous work has shown that with Mel Frequency Cepstral Coefficients~(MFCC) and log Mel-Spectrograms, 1D audio data can be transformed into 2D matrices~\cite{song2021frustration}. This makes it possible to directly apply CNNs, typically from CV, and which have become the mainstream in Computer Audition. In the task of COVID-19 detection, Chang \et \cite{chang2021} studied the performance of classical CNNs pretrained on the FluSense database, collected to track influenza-related indicators, such as cough and sneezes \cite{al2020flusense}. Similarly, Casanova \et \cite{casanova21_interspeech} employed transfer learning from pretrained audio neural networks with different data augmentation techniques.

On the other side, as audio data is inherently a type of temporal sequence \cite{vrysis2017extending}, RNNs~\cite{alvarez2021motivic} and LSTM\cite{deng2020exploiting} have been fully adopted to handle the temporal information in several tasks. For instance, Hassan \et \cite{hassan2020} and Pahar \et \cite{pahar2020} evaluated the role of different audio features as input for LSTM-based classification of COVID-19. Similarly, Yan \et \cite{yan21c_interspeech} introduced the Spatial Attentive ConvLSTM-RNN (SACRNN), able to identify the most valuable features through an embedded temporal attention. Various efforts have also explored more efficient CNNs using residual network approaches and ensembles on audio data~\cite{kumar2020end,vrysis20201d,xu2021investigating}. In particular, Byeonggeun \et \cite{kim2021broadcasted} used a residual broadcast block to retrieve temporal features by averaging the frequency features. Finally, Zhang \et \cite{zhang-8960462} proposed a hierarchical structure called pyramidal temporal pooling~(PTP), which can retrieve temporal information by stacking a global PTP layer on multiple local ones.

\section{Proposed Method}
\label{sec:model}
In this section, we propose Temporal-Oriented ResNet~(TorNet), a modified version of the Broadcasting-residual network~\cite{kim2021broadcasted} tailored to audio data, which we present for COVID-19 recognition. In addition, we also propose an Alternating Broadcast Block (AB Block), which contains several Broadcast Residual Blocks~(BC ResBlock)~\cite{kim2021broadcasted} and combines the temporal 
information to the whole feature map and a convolution layer for a better overview of the features. Finally, we use Frequency-wise Instance Normalisation for better domain generalisation~\cite{Nam2018BatchInstanceNF}.   

\subsection{Broadcast Residual Block}

The original ResNet~\cite{kaiming2015resnet} block is described by $y = x + f(x)$, with $f(x)$ being the residual function, and $x$ and $y$ denoting the input and output features, respectively. 
Normally, $f$ utilises 2D-spatiotemporal features~(\ie 2D convolutions). 
To emphasise temporal features, we exploit 1D-temporal features in addition to 2D ones. To highlight the frequency convolution over all blocks, an auxiliary 2D residual connection is added from 2D features. 
To summarise, the BC-ResBlock can be presented as:
\begin{equation}
\label{equ:bcres}
    y = x + f_{2}(x) +BC(f_{1}(avgpool(f_{2}(x)))).
\end{equation}

In \autoref{equ:bcres}, the 2D feature part $f_{2}$ consists of a 3x1 frequency depth-wise convolution followed by SubSpectral normalisation~\cite{chang2021subspectral}, which splits the input frequency into multiple groups and normalises them separately. To obtain frequency-based temporal features, we apply SubSpectral normalisation instead of Batch Norm. Finally, 2D features are averaged over the frequency dimension.

$f_{1}$ is a combination between a 1x3 temporal convolution with Batch Norm and Swish activation~\cite{rama2017searching} followed by a 1x1 point-wise convolution using a channel dropout rate of $p=0.5$. Thus, the broadcasting operation expands the feature map in $ \mathbb{R}^{1 \times w} $to $ \mathbb{R}^{h \times w} $.

\begin{figure}[t!]
\centering
    \includegraphics[width=0.50\linewidth]{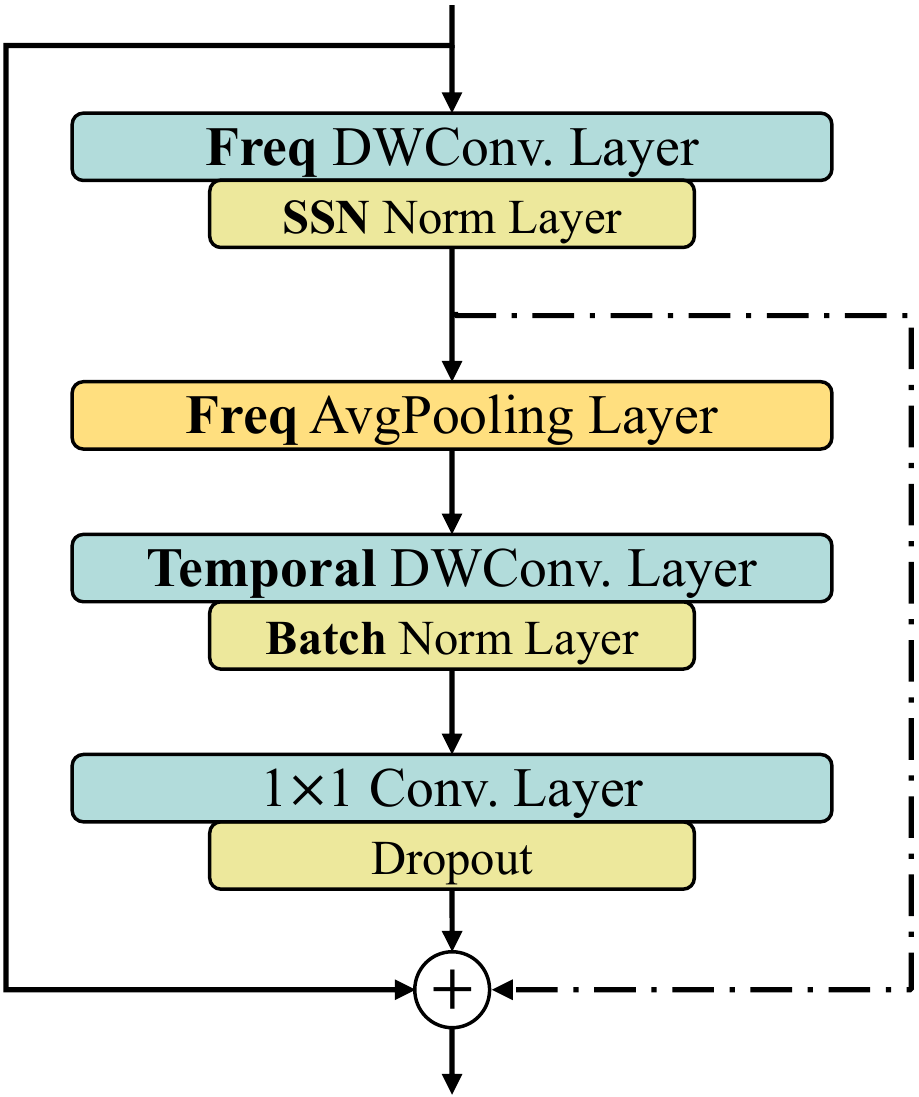}
    \hfill
    \includegraphics[width=0.48\linewidth]{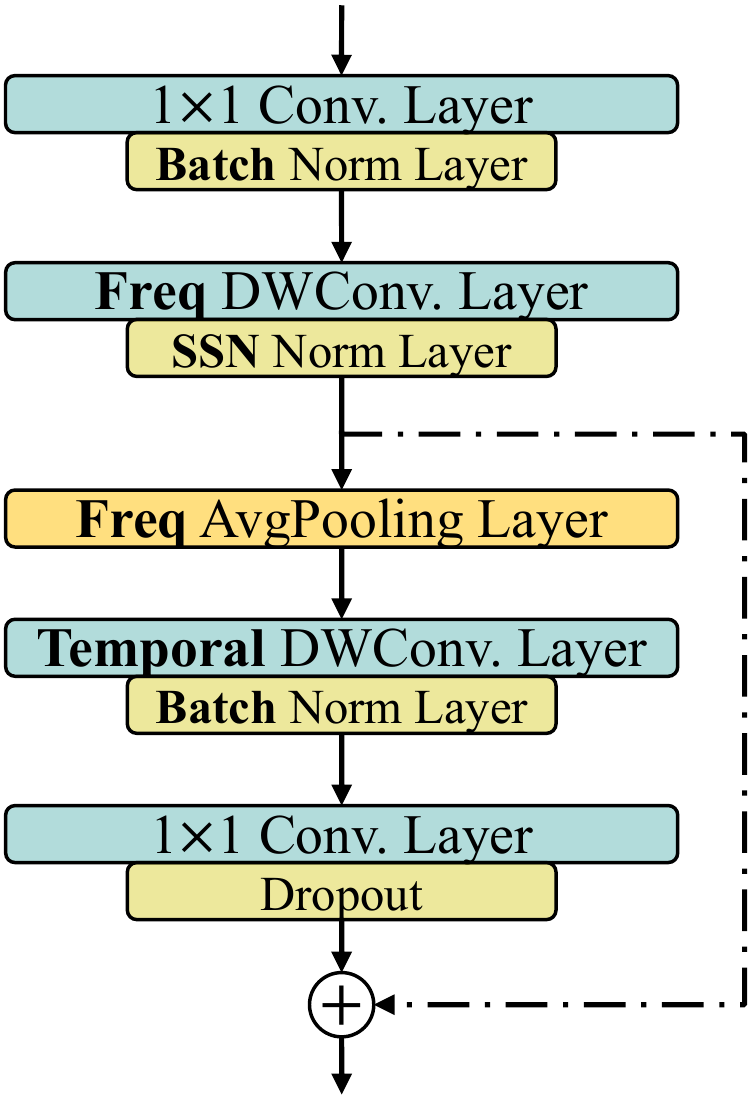}
    \vspace{-0.2cm}
\caption{\textbf{Left: Normal BC ResBlock.} A ResNet structure which firstly compresses the frequency dimension of the feature map by average pooling, and then broadcasts the temporal feature to the original feature map. \textbf{Right: Transition BC Block.} It contains two additional $1\times1$ convolution layers to change the channel number without the identity shortcut.}
\label{fig:bcrb}
\end{figure}

A normal BC ResBlock~(cf.\ left in \autoref{fig:bcrb})~remaps the temporal information to the original feature map, so it has the same input and output dimensions. 
Meanwhile, a transition block~(indicating that the number of input channel and output channels is different)~is used, with the following  modifications:

\begin{enumerate}
    \item When channels do not have the same size, we add a transition block with Batch Norm and ReLU activation;
    \item There is no identity shortcut. 
\end{enumerate}

\subsection{Alternating Broadcast Block}
\begin{figure}[t!]
  \centering
  \includegraphics[width=0.62\linewidth]{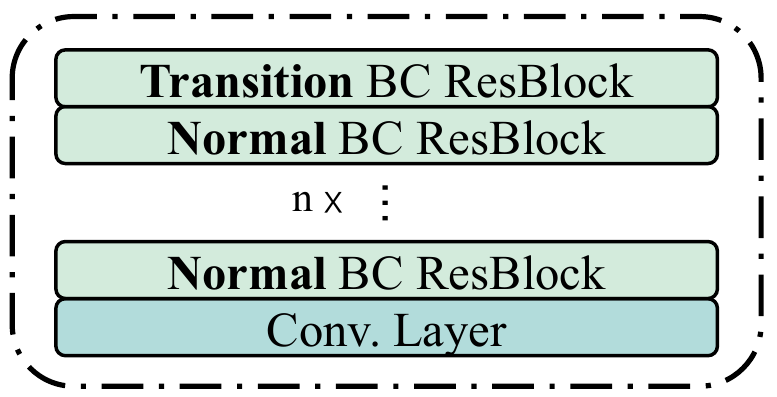}
      \vspace{-0.2cm}
  \caption{The Alternating Broadcast Block~(AB Block).}
  \label{fig:module}
   \vspace{-0.3cm}
\end{figure}

With the BC ResBlock, it is possible to turn the features into a higher dimensionality while broadcasting the temporal information to the whole feature map. As shown in \autoref{fig:module}, we propose a flexible structure of the Alternating Broadcast Block~(AB Block), which mainly contains a set of BC ResBlocks and a convolution layer. The AB Block can be easily widened or deepened, even when facing a large amount of data, by simply adding a larger number of Normal BC ResBlocks. Note that the first BC ResBlock must 
be a Transition BC ResBlock when the number of input and output channels is different. 

As shown in \autoref{fig:bcrb}~(left), average pooling is used before temporal depth-wise~(DW)~convolution, which yields information loss in the frequency dimension~(an inevitable side effect of using the BC ResBlock). In order to reduce the impact of information loss, the last layer of the AB Block is set to a convolution layer, followed by a Batch Norm layer and a ReLU activation layer. The main task of the convolution layer is to capture the global information of the temporal-based feature map, while retaining the local information learnt in the previous layer and projecting them to the higher dimensions of the original inputs. Thus, by using the proposed block, we can achieve enhanced frequency-aware temporal 2D features. 

To achieve a better domain generalisation, we apply Instance Normalisation~\cite{ulyanovVL16}~(IN), an approach that normalises across each channel in each training example. Since IN does not rely on batch information, its implementation is kept the same for both the training and testing phases. We apply IN on the frequency dimension as formulated below:

\begin{equation}
 \label{freqin:main}
  Freq\_IN(x)= \frac{x-\mu_{nf}}{\sqrt{\sigma_{nf}+\epsilon}},
\end{equation}

where
\begin{equation}
 \label{freqin:1}
 \begin{split}
  &\mu_{nf}=\frac{1}{CT}\sum^{C}_{c=1}\sum^{T}_{t=1}x_{ncft},\\
  &\sigma^2_{nf}=\frac{1}{CT}\sum^{C}_{c=1}\sum^{T}_{t=1}(x_{ncft} - \mu_{nf}^2),
 \end{split}
\end{equation}

where $\mu_{nf}, \sigma_{nf} \in \mathbb{R}^{N \times F} $ are mean and standard deviation of the input feature $x \in \mathbb{R}^{N \times C\times F\times T}$, in which $N, C, F, T$ denote the batch size, number of input channel, frequency dimension, and time dimension. $\epsilon$ is a value added to the denominator for numerical stability.

\subsection{Temporal-oriented ResNet~(TorNet)}

\begin{figure}[t!]
  \centering
  \includegraphics[width=0.95\linewidth]{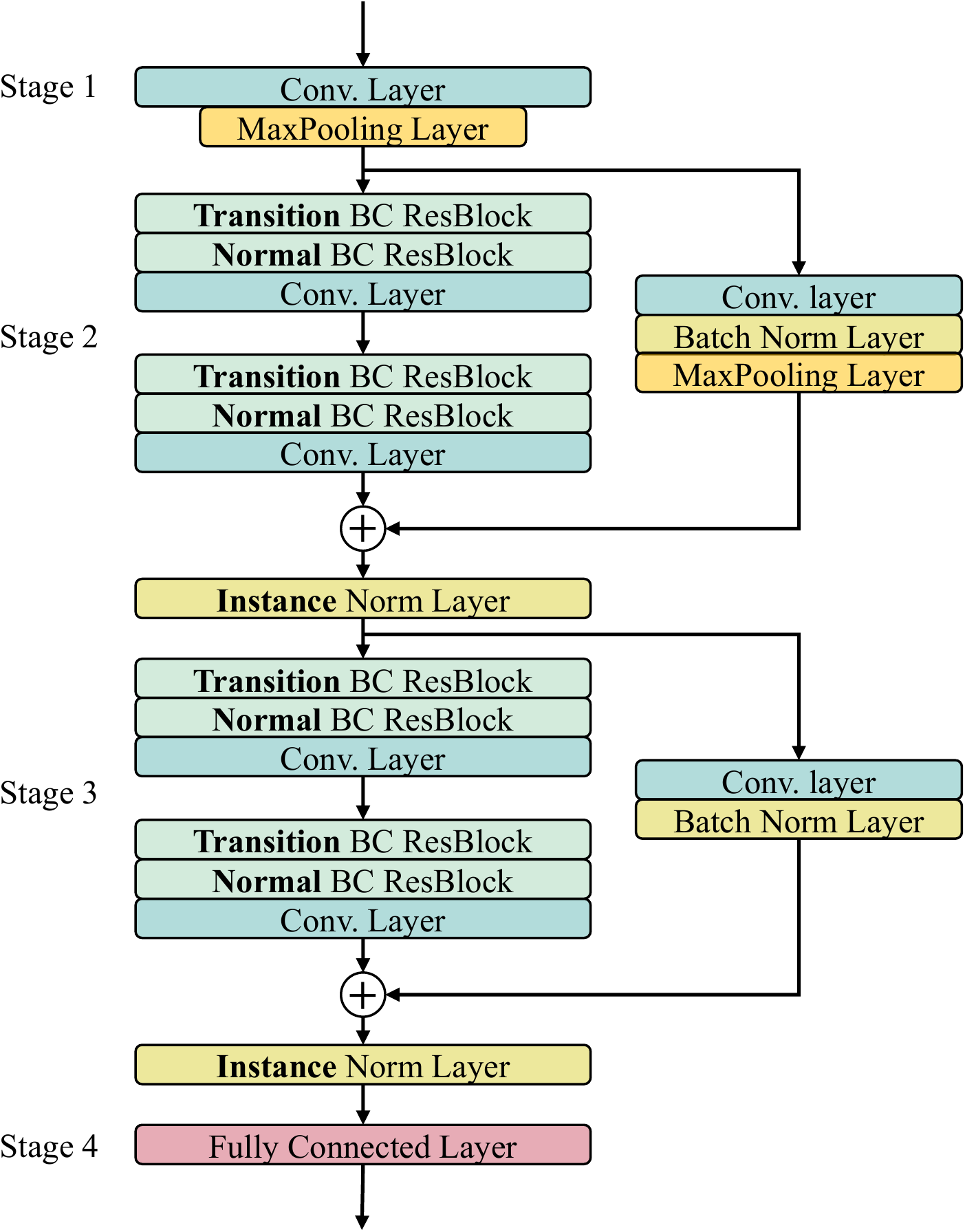}
      \vspace{-0.2cm}
  \caption{TorNet for COVID-19 detection.}
  \label{fig:TorNet}
  \vspace{-0.3cm}
\end{figure}

After exploring several choices and combinations, we design the Temporal-oriented ResNet~(TorNet)~ for the COVID-19 detection task as shown in \autoref{fig:TorNet}. Details on the TorNet structure are given in \autoref{tab:network output}. As shown in \autoref{fig:TorNet}, TorNet contains four main stages. The first stage has a $3\times3$ convolution layer with a $2\times2$ max-pooling layer on the front to downsample both the time and frequency dimensions. The second stage is a typical residual block with two AB Blocks, where every AB Block will double the channel while halving the frequency dimension to get a higher-level embedding. In the residual shortcut, we added a batch norm layer and used maxpooling to control the size of the receptive field. This is followed by an Instance Normalisation layer between stage 2 and stage 3. Stage 3 shares the same structure as stage 2 with minor differences, \ie the number of channels is doubled, and the dimension of the feature map does not change. After the second IN layer, the feature map is turned into a 3D tensor $[batch\_size, time, out\_channel\times N\_mel]$. Finally, two fully connected layers are added as classification layers.

\section{Experiments}
\label{sec:Experiments}
\subsection{Dataset}

In the INTERPSEECH 2021 Computational Paralinguistics Challenge~\cite{schuller2021compare}, the COVID-19 cough sub-challenge~(CCS)~was based on a subset of the crowd-sourced Cambridge COVID-19 Sound database~\cite{brown2020exploring}, whose goal is promoting the developing of systems able to diagnose COVID-19 from audio data. The CCS database consists of $929$ cough recordings~($1.63$\,hours)~from $397$ participants presenting either a positive or negative COVID-19 test. Participants were asked to provide one to three forced coughs in each recording\footnote{https://www.COVID-19-sounds.org/; retrieved 12 March 2022}. All recordings in the CCS database were resampled and converted to $16$\,kHz and mono/16 bits.

The official training, validation, and test sets used in the \textsc{ComparE} challenge are used in all our experiments.

\begin{table}[t!]
\centering
\caption{First 3 stages in TorNet. Each row is a sequence of one module with an input shape of $channel \times frequency \times time$.}
\label{tab:network output}
    \vspace{-0.2cm}
\resizebox{\columnwidth}{!}{
\begin{tabular}{c|c|ccc} 
\toprule
\# & Input & Operator & Stride & Output \\ 
\midrule
\multirow{2}{*}{Stage 1} & 1 $\times$ 40 $\times$ 512 & conv2d 3x3 & 1 & 32 \\
 & 32 $\times$ 40 $\times$ 512 & maxpool 2x2 & 2 & 32 \\
 \midrule
\multirow{2}{*}[-0.5em]{Stage 2} & 32 $\times$ 20 $\times$ 256 & AB Block & ~(2, 1)~ & 64 \\
 & 64 $\times$ 10 $\times$ 256 & AB Block & ~(2, 1)~ & 128 \\
 & 128 $\times$ 5 $\times$ 256 & IN & - & 128 \\
 \midrule
\multirow{2}{*}[-0.5em]{Stage3} & 128 $\times$ 5 $\times$ 256 & AB Block & 1 & 256 \\
 & 256 $\times$ 5 $\times$ 256 & AB Block & 1 & 512 \\
 & 512 $\times$ 5 $\times$ 256 & IN & - & 512 \\
\bottomrule
\end{tabular}
}
\vspace{-0.5cm}
\end{table}

\begin{table*}[t!]
\centering
\caption{Unweighted Average Recall~(UAR), networks' parameters of the overall results and 95\% bootstrap confidence intervals~(CI)~using 1000 samples (with replacement).  The best result of our experiments is marked in \textbf{bold}, and the second best one is \underline{underlined}.}
\label{tab:results}
\scalebox{0.9}{
\begin{tabular}{l|rrl} 
\toprule
\multicolumn{1}{c|}{Method} & \multicolumn{1}{c}{\# Param(M)} & UAR~(\%)~ & CI on Test(\%) \\ 
\midrule
End2You\cite{schuller2021compare} & - & 64.7 & 56.2 - 73.5 \\
Fusion\cite{schuller2021compare}~(official baseline) & - & 73.9 & 66.0 - 82.6 \\
CNN14 \cite{casanova21_interspeech} & 79.67 & 75.9 & - \\
The Vision Transformer (ViT)\cite{illium21_interspeech}  & - & 72.0 & - \\ 
\midrule
ResNet-10~(baseline) & 5.12 & 66.9 & 57.7 - 71.8 \\
ResNet-10+LSTM & 6.17 & 66.7 & 64.2 - 81.6 \\
ResNet-10+LSTM+attention & 6.23 & \textbf{70.5} & 63.9 - 80.1 \\
ResNet-10+Transformer & 58.90 & \underline{68.0} & 63.3 - 79.6 \\ 
\midrule
TorNet (AB Block without last conv) & \textbf{1.32} & 65.5 & 60.1 - 77.6 \\
TorNet (only Transition Block) & \underline{4.09} & 69.4 & 58.5 - 72.6 \\
TorNet    without InstanceNorm & 4.46 & \underline{70.2} & 59.4 - 72.8 \\
TorNet with      InstanceNorm & 4.46 & \textbf{72.2} & 71.5 - 88.6 \\
\bottomrule
\end{tabular}}
\vspace{-0.5cm}
\end{table*}

\subsection{Experiment settings}
For data pre-processing, we standardise the length of the audio data to 10 seconds. The shorter samples are repeated until they match the target length. As input features, we use 40-dimensional log Mel-Spectrograms with a 64\,ms window length and a 16\,ms frame shift. We also extract deltas and delta-delta of log Mel-Spectrograms and concatenate them as input features. At last, the size of features for TorNet is $[batch\_size, 3, 40, 512]$. For all models, we use the Adam optimiser with an epsilon value of $10^{-8}$, a mini-batch size of $16$, and a learning rate of $10^{-5}$. As indicated by \citet{chang2021subspectral}, the sub-bands of SSN in AB Blocks were all set to $5$ and the dropout rate was always $p=0.1$ except for the last layer, where it was set to $p=0.5$. We also tried data augmentation~(mixup, Spec augmentation)~methods, but there was no noticeable performance gain in our task, thus omitting them for brevity.
All the models were developed on Pytorch 1.8.1 and trained on a single Nvidia RTX 3090 GPU.

\subsection{Proposed Experiments}
To verify the efficiency of the proposed TorNet, as well as the effectiveness of the temporal features, we developed four additional ResNet-based methods for comparison. These four methods are as follows:
\begin{itemize}
    \item ResNet-10: has an identical structure as TorNet but uses a convolution layer and maxpooling to control the size of the feature map. It is used as a baseline model.
    \item ResNet-10 + LSTM: introduces a layer of a standard LSTM network at the output of the ResNet-10.
    \item ResNet-10 + LSTM + Attention: adds a 4-head multihead attention module to extract more centralised temporal-frequency feature maps.
    \item ResNet-10 + Transformer: adds 2 transformer encoder layers for locating and re-extracting the most relevant features of the audio segments.
\end{itemize}

\section{Results and Discussion}
\label{sec:results}

Our experimental results obtained on the binary task of COVID-19 detection are presented in \autoref{tab:results}. The upper part of \autoref{tab:results} displays the results from previous works on the CSS dataset. The middle part shows the results obtained from the four additional ResNet-based methods presented for comparison. Finally, the results for the series of experiments with same network hyperparameters and different modules are given in the lower part. Unweighted Average Recall~(UAR)~is reported as the evaluation metric. For each method, the results on the test set are obtained by using the model achieving the highest UAR on the validation set.

Our proposed TorNet, based on the combination of AB Block and residual learning's results, reached up to 72.2\% 
UAR. The results show that all experiment results on the Tornet outperform the official baseline~(End2You 64.7\%)~but still lag behind other approaches.
Unlike the herein presented one, \citet{casanova21_interspeech} achieved 75.9\% UAR based on a large-scale transfer learning model. 
Their CNN14 model is pre-trained on Audioset, which means a longer training time and higher computational effort. 
Indeed, the parameters of CNN14~(79.67 millions)~are almost 18 times more than the number of parameters compared to TorNet ~(4.46 millions) -- thus showing that TorNet has a higher computational efficiency. 
Similarly, the baseline fusion framework for the CCS Sub-Challenge fuses multiple best models to obtain the final results~(73.9\% UAR), which also results in a far higher computational complexity than our TorNet.
Overall, this shows that TorNet can achieve competitive performance without pre-training or fusion while also using far lower computational resources.

In comparison to the official `standard' baseline~(End2You), our baseline results show that the ResNet structure still has good robustness for audio data~(ResNet-10 66.9\%). 
Meanwhile, we added LSTM and transformer structures for extracting temporal information after ResNet-10. The results show that the extraction of temporal information can  improve the final UAR results, as shown by the  combination of ResNet + LSTM~(cf.\ 70.5\%,  in the bold, in the  middle part of \autoref{tab:results}); by this, outperforming all other ResNet-based methods.

These experiments demonstrate the importance of utilising temporal context. However, these architectures dramatically increase the computational complexity while still not reaching the performance of TorNet. This demonstrates how accounting for the temporal nature of audio inside the intermediate layers of a DNN --like TorNet does-- is superior to doing it only towards the deeper part of the network.

Since our goal is to investigate to which extent it is possible to model temporal information while improving computational efficiency with DNNs, we also set up four comparison experiments based on TorNet. In these, we keep all training parameters consistent in order to assess the impact of different modules, \ie the use of a convolution layer in the AB Block, the Normal BC ResBlock, and Instance Normalisation on the overall performance of TorNet.

The lower part of \cref{tab:results} contains an ablation study of the components introduced in this work.
TorNet without the last convolution layer in the AB block achieves only 65.5\% 
UAR, while when convolution layers are introduced, there is a performance improvement of nearly 5.0\%~(cf.\ 70.2\%, underlined in the lower part of \autoref{tab:results}). 
This is because in each AB Block, the BC ResBlock has the ability to broadcast the temporal features to the original feature map, but loses a portion of the frequency features. 
By introducing an extra convolution layer, we eliminate the influence that this loss of granular details entails, thus obtaining a better overview for the feature map, which results in a sizable performance increase. 
At the same time, the IN layer is introduced in the frequency dimension for better domain generalisation, leading to a performance improvement of approximately 2\% in the same training environment: from 70.2\% (without IN) to 72.2\%~(with IN).

\section{Conclusion}
\label{sec:conclusion}
In this work, we proposed an AB Block that can efficiently exploit the temporal information in audio sequences. It contains multiple BC ResBlock as well as a convolution layer to capture the temporal-enhanced feature. Based on the AB Block with residual learning, we proposed a flexible, lightweight, and time-oriented network -- TorNet. TorNet has a typical ResNet structure, but we replace the convolution module with the AB Block. Competitive results highlight the high computational efficiency and robustness of TorNet, a promising architecture that offers new insights for the detection of COVID-19.

Future work could be targeted towards the application of TorNet in other domains, such as speech emotion recognition or acoustic scene classification. 


\section{\refname}
 \printbibliography[heading=none]

\end{document}